\documentclass{sig-alternate-10pt}
\usepackage{url}
\usepackage{graphicx}
\usepackage{xspace}
\usepackage{listings}
\usepackage{fancyvrb}

\def\true{true}

\def\notdraft{true}

\newcommand{\eg}{e.g.,\xspace}
\newcommand{\ie}{i.e.,\xspace}
\newcommand{\etal}{et al.\xspace}
\newcommand{\naive}{na\"{\i}ve\xspace}
\newcommand{\sami}{SAAMI\xspace}
\newcommand{\pktin}{\texttt{OFPacketIn}\xspace}
\newcommand{\pktout}{\texttt{OFPacketOut}\xspace}
\newcommand{\ping}{\texttt{ping}\xspace}
\newcommand{\tr}{\texttt{traceroute}\xspace}
\newcommand{\of}{OpenFlow\xspace}
\newcommand{\ovs}{Open vSwitch\xspace}

\begin{document}

\title{SDN as Active Measurement Infrastructure}

\numberofauthors{2}
\author{
 \alignauthor Erik Rye\\
 \affaddr{US Naval Academy}\\
 \email{rye@usna.edu}
 \alignauthor Robert Beverly \\
 \affaddr{Naval Postgraduate School}\\
 \email{rbeverly@nps.edu}
}

\maketitle
\thispagestyle{empty}

\begin{abstract}
Active measurements are integral to the operation and management of
networks, and invaluable to supporting empirical network research.
Unfortunately, it is often
cost-prohibitive and logistically difficult to widely deploy
measurement nodes, especially in the core.  In this work, we consider
the feasibility of tightly integrating measurement \emph{within}
the infrastructure by using
Software Defined Networks (SDNs).
We introduce 
``SDN as Active Measurement Infrastructure'' (\sami) to enable
measurements to originate
from any location where SDN is deployed, removing the need for
dedicated measurement nodes and increasing vantage point 
diversity.  
We implement \ping and \tr 
using \sami, as well as a proof-of-concept custom
measurement protocol to demonstrate the power and ease of \sami's open
framework.  Via a 
large-scale measurement campaign using SDN switches
as vantage points, we show that \sami is accurate, scalable,
and extensible.
\end{abstract}

\section{Introduction}

Software Defined Networking (SDN) has emerged as a powerful
architectural paradigm, enabling innovations in network
virtualization, provisioning, verification, and security
(e.g., \cite{foster2013languages, gupta2015sdx, jafarian2012openflow}).
Within the context of measurement, SDNs are commonly instrumented to monitor
network utilization and can quickly modify their forwarding behavior
in response to dynamic workloads \cite{tootoonchian2010opentm}.  In
contrast to such passive
measurements (e.g., packet or flow-level statistics, heavy-hitters, or
anomaly detectors), this work considers the feasibility of performing
\emph{active} network measurement using SDNs.

Intuitively, SDNs provide the basic building blocks for
facilitating programmable active measurement vantage points.  Via the
standardized \of protocol \cite{mckeown2008openflow}, SDN controllers
can generate any arbitrary packet (\eg measurement probes) and instruct 
a switch to emit the packet
out a specified interface.  Similarly, controllers can instantiate
fine-grained flow rules that match measurement probe responses. 
Controllers may then perform arbitrary computation over probe
responses as they arrive encapsulated in an \of message from the
switch.
In this work, we
introduce and advocate for an architecture using these primitives
which
we term ``SDN as Active Measurement Infrastructure''
(\sami).  

While active measurements today are performed using end-hosts, we
believe \sami provides compelling
advantages.  First, tighter integration of active measurements within
the network allows operators, administrators, and researchers to place
measurements anywhere an SDN switch exists -- without the traditional
cost of configuring, certifying, securing, installing, powering, managing and maintaining a dedicated measurement
host.  Often, installing a measurement host within the core or edge of
the network presents an insurmountable administrative or policy hurdle, or is
not physically possible (due to space, availability and expense of
consuming an interface on the router, etc).  By leveraging existing production equipment,
\sami lowers deployment and vantage point diversity barriers.

Second, placing active measurements within SDN more closely couples
the ability to programatically enable action to be taken in response
to measurement results.  In a similar fashion to using passive
measurements to drive the behavior of the SDN
data plane, active measurements can provide actionable information for
network reconfiguration and adaptability.

Third, \sami follows the
philosophy of inexpensive commodity hardware, centralized control,
and compositional network architecture.  \sami uses the 
standardized \of API to drive commodity switches, thereby removing the traditional need
to develop, configure, and deploy custom APIs for requesting
measurement tasks and receiving measurement results.  As we will show, using \of
prevents over-specialization of the measurement platform, making it
easy to extend to new and unanticipated measurement tasks.

Toward the goal of converging on a standards-based strategy for 
active measurement platforms for operators and researchers, we 
investigate the feasibility and limitations of \sami.  
Our primary contributions include:
\begin{enumerate}
 \addtolength{\itemsep}{-0.5\baselineskip}
 \item The \sami architectural vision to converge active
       measurement facilities into commodity hardware using 
       standardized protocols.  
 \item A large-scale Internet measurement study using \sami to source
       active probes from commodity SDN switches, and a
       corresponding analysis of the resulting fidelity of
       round-trip latency measurements.
 \item Implementation of a new measurement protocol using the \sami
       primitives as a demonstration of the ability to easily and rapidly 
       innovate and deploy measurements within the \sami framework. 
 \item A fundamentally different approach to Internet measurement, in which core
       nodes (in addition to edge systems) participate as vantage points for active
       measurements, potentially in reaction to events in the traffic streams
       they observe. 
\end{enumerate}


The remainder of this paper is organized as follows.  Section
\ref{sec:implementation} describes \sami, while Section
\ref{sec:results} details results and accuracy from using \sami in 
real-world large-scale Internet measurement experiments.  We discuss
related work in Section~\ref{sec:related}, and conclude with a
discussion of deployment scenarios and suggestions for future work in
Section~\ref{sec:conclusions}.

\section{Implementation}
\label{sec:implementation}

Active measurement is widely used for network management and
debugging.  Canonical examples include the \ping and \tr
utilities that provide reachability, round-trip time, and
forward path information.  Not only are such tools used for
troubleshooting, they are integral to the operation of large providers,
content distribution networks, and at-scale services.

Myriad other active measurement techniques and tools have been
developed, for instance host and service fingerprinting
\cite{lyon2009nmap}, capacity and bandwidth estimation
\cite{dovrolis2001packet}, censorship detection
\cite{burnett2015encore}, network neutrality
\cite{dischinger2010glasnost}, residential broadband performance
\cite{bauer2010understanding}, and congestion localization
\cite{luckie2014challenges}, to name only a few.

It is well-known that the accuracy and generality of inferences from
these active measurements can depend strongly on where in the network the
measurements are performed.  For instance, censorship may only affect
nodes behind a certain middlebox, congestion may affect only a single
network, or a BGP hijack event may affect reachability for only a
subset of vantage points.  To this end, the research community has
developed several active measurement platforms consisting of varying
(but relatively small) numbers of nodes distributed (in an ad-hoc
fashion) across the Internet, e.g., Archipelago \cite{hyun2015caida},
Bismark \cite{sundaresan2014bismark}, Dasu \cite{sanchez2013dasu}, and
Atlas \cite{ripe2010ripe} to name a few.  Unfortunately, these
platforms are frequently: i) designed for a specialized task; ii)
under-powered (either in terms of compute, memory, or network
abilities); and iii) lacking in network or geographic diversity.

Rather than the current fractured environment of incompatible network
measurement platforms, abilities, APIs, and output formats, we
consider the feasibility of performing active measurement 
using existing SDN standards and capabilities.
Our vision that by using SDN, the measurement and
operational community can utilize a standard and open API to avoid
over-specialization, lower deployment barriers, and facilitate vantage
points otherwise not possible.

While significant prior work utilizes the \emph{passive} network
measurement capabilities afforded by SDNs, e.g., \cite{6888899,
tootoonchian2010opentm, narayanahardware, jose2011online, 180293} and
novel methods to gather measurements from SDNs, e.g.,
\cite{Yu:2013:FMN:2482362.2482367}, our work seeks to
understand the feasibility of SDN for \emph{active} measurement --
i.e., where the SDN switch generates specialized measurement probes and
gathers their responses.  

\begin{figure}[t]
  \centering
  \resizebox{0.4\columnwidth}{!}{\includegraphics{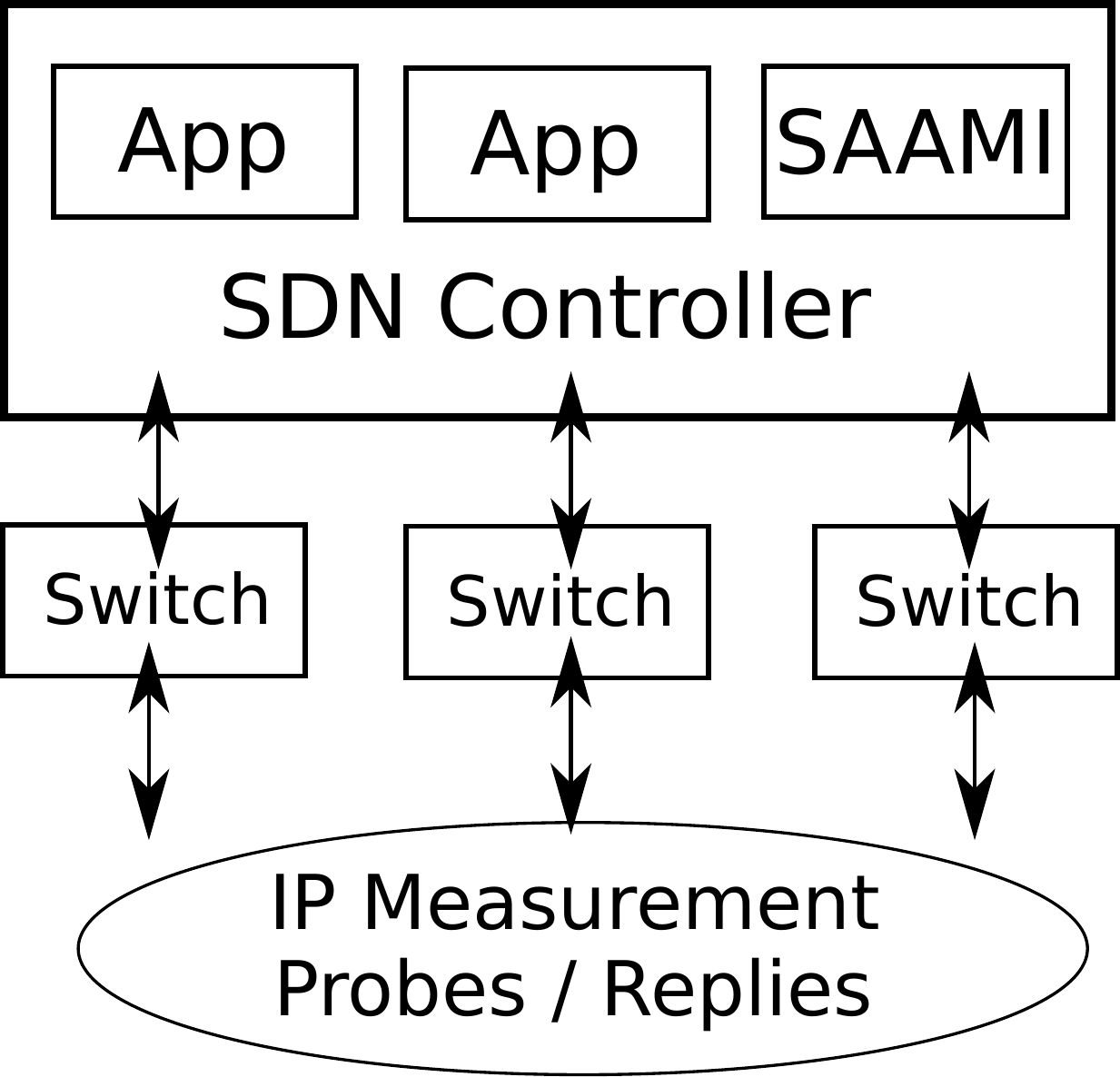}}
  \caption{\sami architecture: \sami acts as
  an abstraction layer between various active IP measurement tasks
  and the underlying SDN.}
  \label{fig:arch}
\end{figure}

Figure~\ref{fig:arch} illustrates the high-level architectural view of
\sami within an SDN.  We envision \sami co-existing with other
applications on an SDN controller that is responsible for one or more
SDN switches within a provider or enterprise network.  

Our
implementation is based on the popular open-source Ryu~\cite{ryu2013framework} controller.  In
response to measurement tasks (\S\ref{sec:api})
\sami instructs the SDN controller to send various
\of~\cite{mckeown2008openflow} messages to the SDN switch(es) in
order to induce active IP measurement probes.  In particular, we
utilize the \pktout message to instruct a switch to send a
particular packet. 
Further, \sami installs
flow table rules such that it can receive and process measurement
replies;
we rely on \pktin messages from the
switch that encapsulate data-plane probe packets matching particular criteria.
In this fashion, \sami acts as the abstraction layer between
measurement tasks and the SDN.   

\subsection{Calibration}
\label{sec:calibration}

We note that many measurement tasks, including \ping and \tr, require accurate timing
information, \eg for round-trip time (RTT) latency estimation.
\sami must thus address two timing challenges related to using
commodity switches and the \of protocol: approximating the time
at which measurement packets are sent and received.  With respect to probe
transmission, both the delay in sending the \of instruction (the
``\pktout'') from the controller to the switch, and the switch's
processing delay in executing that instruction, contribute potentially variable
latency.  Further, while \of provides a mechanism to forward
probe responses from the switch to the controller (the ``\pktin''), there is no 
standardized way to obtain the time when the packet was received at
the switch.  We must therefore estimate these values via an
assessment of the RTT between the controller and the switch, and
via empirical analysis of commodity SDN switch behavior.  While
it is possible to design specialized hardware and implement
API changes to address these sources of inaccuracy, our goal in this
work to determine the feasibility of \emph{current} SDN
hardware and software implementations to support \sami.

\subsubsection{Switch Processing Delays}

\begin{figure}[t]
  \centering
  \resizebox{0.7\columnwidth}{!}{\includegraphics{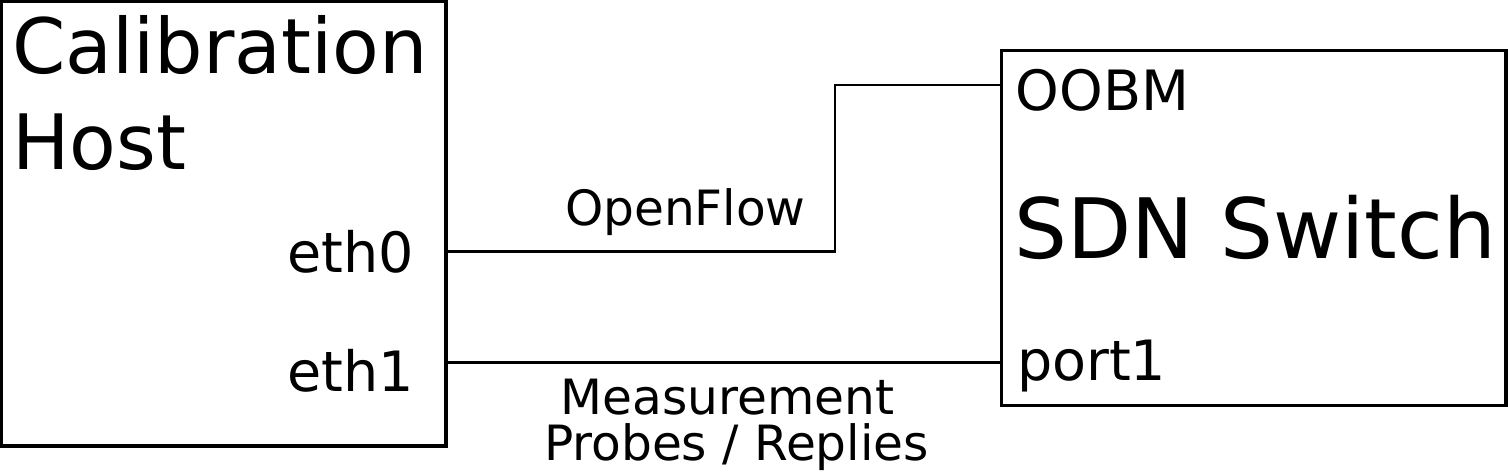}}
  \caption{Calibration: Isolated testbed to empirically measure
  delay between issuing measurement probes and receiving responses
  via \of with a commodity SDN switch.}
  \label{fig:calibration}
\end{figure}

Let $R(X)$ be the time at which some packet $X$ is received at a
switch port and $T(X)$ be the time at which some packet $X$ is
emitted from a switch port.

\begin{itemize}
  \item Assume a switch receives a probe within a \pktout at
        $R(pktout)$.  To determine the switch's delay
        is emitting the probe, we compute: $T(probe) - R(pktout)$. 
  \item Further, we wish to determine whether any output reordering
        occurs, \ie $\exists i \text{ s.t.\ } T(probe_{i+1}) <
        T(probe_i)$ and $R(pktout_i) < R(pktout_{i+1})$.
  \item Assume a switch receives a response packet that matches
        a flow rule.  The switch's delay in generating
        the \pktin message is: $T(pktin) - R(response)$.
\end{itemize}

To isolate sources of delay and calibrate our \sami measurements, we
create an isolated testbed using a Linux machine with two physical
Ethernet interfaces as shown in Figure~\ref{fig:calibration}.  One 
Ethernet 
directly connects to the Out-of-Band Management
(OOBM) port on a commodity commercial SDN switch, while the other
Ethernet is connected to one of the switch's data-plane interfaces.
In our experiments, we use a commodity
HP2920 commercial switch running \of 1.3.  We then perform a
packet capture from the Linux machine to time \of packets in
relation to dataplane packets (either generated or received).  Because
the packet capture is performed on two interfaces of the same physical
machine, time is synchronized.
Since the host and the
SDN switch are directly connected via a $<1$ meter cable, propagation delay
is a negligible component of the measured delay.

\begin{figure}[t]
  \centering
  \resizebox{0.8\columnwidth}{!}{\includegraphics{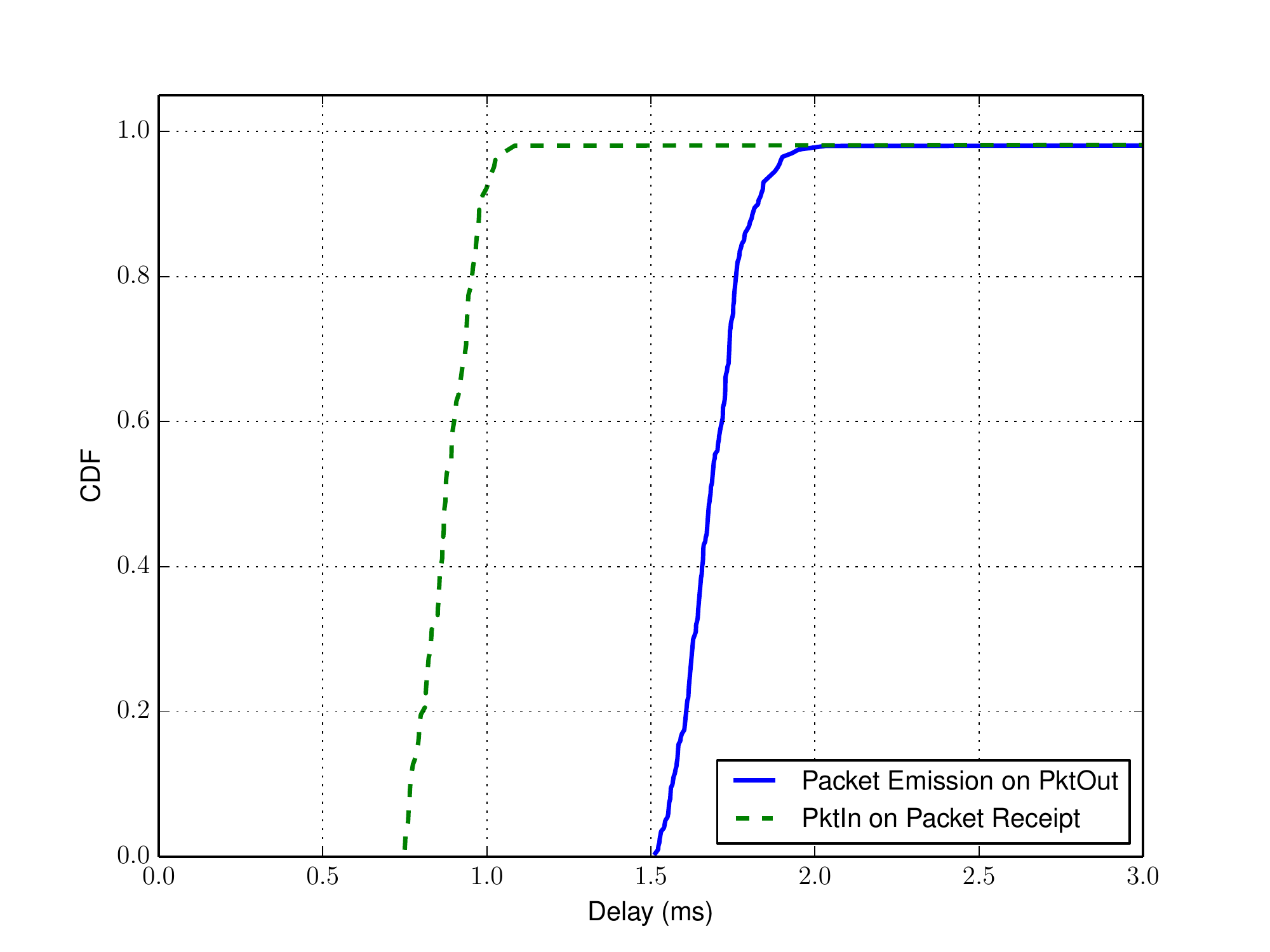}}
  \caption{Packet emission delay for a commodity SDN switch in
   response to \pktout \of messages.}
  \label{fig:calibpktout}
\end{figure}

We send 100 \pktout messages to the switch, and measure the switch's
delay in emitting the corresponding packet.
Figure~\ref{fig:calibpktout} displays the cumulative fraction of
delays $T(probe_i) - R(pktout_i)$ over the 100 \pktout requests.  We see that 97\%
have a delay between 1.5 and 2.0ms.  Two \pktout messages require
$\approx 23$ms, while the switch took approximately 50ms before
emitting one of the packets.  Overall, the delay
is both small and tightly bounded.  Further, we observe no packet
reordering.

Similarly, we evaluate the switch's delay in emitting a \pktin message
in response to receiving a packet that matches a flow rule,
$T(pktin_i) - R(response_i)$.  We see a qualitatively similarly shaped
distribution in Figure~\ref{fig:calibpktout}, however 95\% of the
\pktin messages are generated in 1.0ms or less.

\subsubsection{Bundled Messages}

We observe instances of multiple \pktout
messages bundled into single TCP segments by the \sami controller. 
Such effects are due to operating systems and their
corresponding TCP stack implementations.
Quantifying the effect that bundling has on time-sensitive
measurements is therefore important for \eg RTT estimation. Thus,
we must assess the variation in time for packets to be emitted by an
SDN switch, whether arriving at the switch as \pktout messages in separate TCP
segments, or bundled with other \pktout messages in a single TCP segment.

\begin{figure}[t]
  \centering
  \resizebox{0.8\columnwidth}{!}{\includegraphics{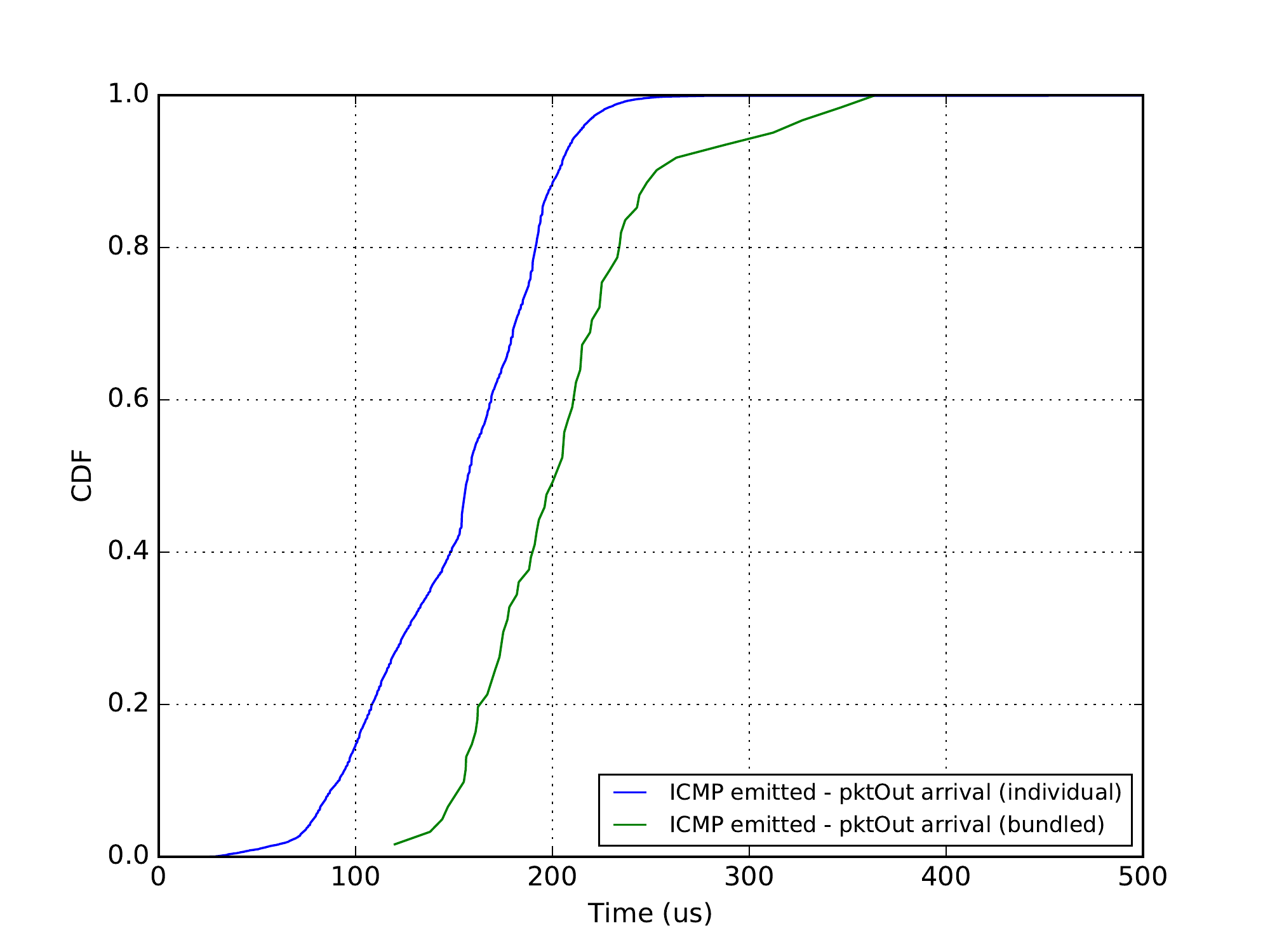}}
  \caption{Calibration: \ovs delay for ``bundled'' \pktout messages
    vs.\ individual \pktout messages }
  \label{fig:blobSwitchTimeDiffs}
\end{figure}

In order to measure this effect, we use \pktout messages to instruct
the switch to emit ICMP packets.  We calculate the difference in time between an
ICMP Echo Request leaving the SDN switch and its corresponding \pktout message arriving at the
switch, accounting for \pktout messages in individual segments and multiple
\pktout messages contained in a single segment separately. We sent 15,000
\pktout probe requests with 5 probes apiece to a host running \ovs
\cite{pfaff2009extending} acting as the SDN
switch. Of the 75,000 probes generated, only 61 probes are sent as 
\pktout messages contained in a TCP segment with other \pktout messages. The
time difference between ICMP Echo Requests exiting the switch and the time
the \pktout message entered the switch is characterized in
Figure~\ref{fig:blobSwitchTimeDiffs}. On average, the overall switch delay is
approximately 10\% higher for probes sent in multiple-\pktout-containing
segments -- a 206 $\mu s$ delay was incurred by multipart messages, whereas
individually sent \pktout pings had only a 189 $\mu s$ mean delay.

\subsubsection{Controller to Switch RTT Estimation}
\label{sec:c-sEst}

The \sami controller measures the RTT between
controller and associated SDN switch by issuing and timestamping an
\texttt{OFEchoRequest} message, a built-in \of message type that is
used by default as a ``heartbeat'' between the controller and switch. When the
corresponding \texttt{OFEchoReply} returns from the switch, the controller
timestamps this reply, allowing for the calculation of the RTT between the
controller and switch $RTT_{C-S}$. For our \of implementations of
\texttt{ping} (Section~\ref{sec:ping}) and \texttt{traceroute}
(Section~\ref{sec:traceroute}) we estimate $RTT_{C-S}$ for each target
separately before generating probes. In order to account for isolated, drastic
changes in controller-to-switch latencies, we keep an exponential moving weighted
average of $RTT_{C-S}$ times for use in true RTT estimation calculations.

Note that performing the controller to switch latency estimation
\emph{within} the \of protocol, as opposed to, e.g., using
a simple ICMP echo, provides a more reliable approximation of the
delay (and processing) incurred by \of messages.


\subsection{Configuration}
\label{sec:arp}

For experiments in which probes must eventually return to the switch that
emitted them (\eg \ping and \tr), these probes must use a source
address that is routed to or through the
switch that generated the probe\footnote{We discuss more complicated
scenarios where a different switch in the network receives probe
responses in \S\ref{sec:conclusions}.}.  Thus, \sami must know what source IP 
address to use when generating \pktout messages and installing flow
rules.  While this requires specific addressing information to be
known, such knowledge is integral to SDN controllers.  

In our testing, we chose an unused address on a subnet that is routed
to a network on which the SDN switch is connected.  Because it was not
possible to assign an IP address to the commodity SDN switch port, 
\sami deliberately generates
gratuitous ARPs.  This allows the router to which our SDN switch
is connected to pre-populate its ARP cache and prevent any additional
delay. 
We expect that generating these ARPs will not be
necessary in other deployments where SDN switches act as layer-3
forwarders.

Thus, for correct operation, \sami must at a minimum 
be configured with: i) the IP and MAC addresses to use when sourcing
measurement probes; and ii) the IP and MAC address of the next hop such
that the probe is properly forwarded.  As is typical in SDN
installations, the switches are configured with the IP address of the
controller, and \sami listens for incoming \of connections.


\subsection{Measurement API}
\label{sec:api}

Following our general philosophy of leveraging existing protocols,
\sami utilizes HTTP and JSON as its measurement API with which
the consumers of measurement tasks interface.  In this way, \sami
acts as the interface between high-level measurement tasks, and
the marshaling of \of messages within the SDN plane.

The \sami controller therefore runs an HTTP server that listens for incoming
measurement instructions.  We leverage HTTP as a standard RPC-like
mechanism with the ability to easily support encryption, integrity,
and authentication.

As a concrete example, we detail here the REST API calls necessary for our \ping
implementation described in Section~\ref{sec:ping}. Many common measurement
tasks may leverage or require the emission of ICMP Echo Request packets, such as
RTT estimation or to elicit responses containing IP Identifier values. To enable
these experiments, the \sami controller has a defined ``probe URL'', a URL that
when requested with the required parameters included in a JSON array via an HTTP
PUT, will emit Echo Requests destined for a particular target. For instance, 
the controller uses URL \texttt{http://example.com:8080/ping/} as the Echo
Request-emitting resource locator (by default, Ryu listens for API calls on port
8080). Our \ping implementation requires 
a JSON array be sent with the HTTP PUT request containing three keywords
extending the JSON schema:
i) \texttt{tgt}, the IP address of the target; ii) \texttt{num}, the number of Echo Requests
to be emitted; and iii) \texttt{payload}, an optional payload for inclusion in the Echo
Requests \sami emits from the SDN switch. Upon receipt of the PUT
request, the \sami controller parses the JSON
array and passes the fields to the method responsible for creating the ICMP Echo
Requests, encapsulating them in IP datagrams and Ethernet frames, and delivering
\pktout messages to the SDN switch. Switch for packet emission
and output port selection are specified in \pktout message data fields, ensuring
the correct switch emits the packet from the proper output interface.
The use of the
REST API creates an extensible measurement framework -- in our \ping example, it
is trivial to add an additional optional source IP address parameter for Echo
Request generation or to define a default TTL for the emitted datagrams -- that
is easily automated using scripts (\eg with \texttt{curl} or
\texttt{wget}). Furthermore, the API
enables the experimenter to operate independently from the measurement platform,
without needing to log into the controller to start or stop measurements.

Experiment output is retrieved via REST API calls; we obtain state
maintained by the \sami controller again through the use of HTTP GET requests to
predefined URLs. For example, an HTTP GET request to
\texttt{http://example.com:8080/ping/dump} returns a JSON array containing the
packet emission and arrival times associated with each ICMP ID and Sequence
Numbers sent in our \ping implementation, using the ICMP ID as the JSON schema
keyword. 
ICMP Echo Requests that receive no corresponding Echo Reply simply contain a
null entry in the response timestamp position in its corresponding field,
allowing unresponsive destination hosts to be treated differently than
responsive destination addresses. This raw data can then be manipulated by the
experimenter according to their own needs. A practical application using this
data (estimating RTTs) is demonstrated in Section~\ref{sec:ping}.

\begin{figure}[t]
  \centering
  \resizebox{\columnwidth}{!}{\includegraphics{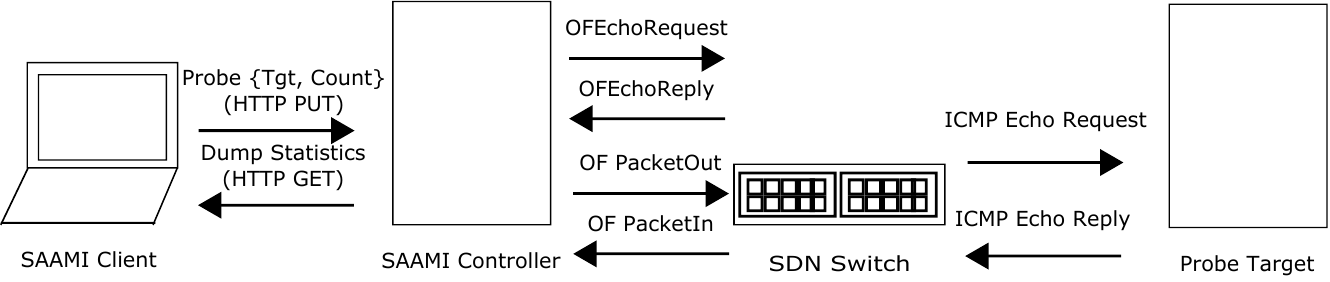}}
  \caption{Depiction of \sami \ping implementation.}
  \label{fig:ping_arch}
\end{figure}

Figure~\ref{fig:ping_arch} is a high-level view of \sami's \ping implementation.


\subsection{Common Measurement Tasks}

\subsubsection{Ping}
\label{sec:ping}

\sami implements the \ping network utility through an HTTP PUT message
to the \sami controller from a \sami client.   
This PUT message includes a JSON
array with target IP address, count and payload fields. 
In our implementation,
the \sami client is a trivial 12 line Python program, illustrating the
ease with which it is possible to develop within the \sami framework.

Upon receipt of the \ping
request via the PUT, the controller first sends an ARP reply to the router
containing the MAC and IP addresses of the controller's interface to the switch
as described in \S\ref{sec:arp}.  This gratuitous ARP reply prevents any
ARP cache expiration; 
without this gratuitous ARP
reply, we risk \sami needing to reply to ARP requests and
thereby negatively affecting time-sensitive RTT estimation.

Next, \sami initiates RTT estimation to the switch as outlined in
Section~\ref{sec:c-sEst}.
The controller then creates an ICMP Echo Request destined for the target IP
address, encapsulates it in an Ethernet frame addressed to the next hop's
MAC address, and emits an \pktout message to an SDN switch. Concurrently, the
controller creates a map between the ICMP Identifier field value used and target
IP address to maintain state of in-flight probes. \sami initially sets the ICMP
Sequence Number field to 0 for the first ICMP Echo Request. The payload of our
Echo Request is empty, but can be specified by the requester in the JSON array
sent with the HTTP PUT.

\sami timestamps the packet and delivers a \pktout message to the SDN switch,
which emits the ICMP Echo Request out of the switch port specified as an
argument to the \texttt{OFPActionOutput()} method. \sami then increments the
ICMP Sequence Number field value and repeats until the requested number of ICMP
Echo Requests have been delivered to the switch via \pktout messages. For each
subsequent ping target, \sami increments the ICMP Identifier field value,
establishing a linkage between ICMP Identifier values and ping destination IP
addresses. 

Concurrently, \sami is delivered ICMP Echo Replies for the probes it has
transmitted by the switch via \pktin messages, due to the installation of a
flow rule during \sami initialization directing the switch to forward these to
the controller.
When a probe reply is received \sami notes the time the ICMP Echo Replies were
received by the controller
and stores this value. For a given target $Tgt$, we can
therefore calculate $RTT_{C-Tgt}$, the total time elapsed between
\pktout transmission from \sami to the switch and
\pktin messages received by \sami from the switch for each ICMP
Sequence Number. $RTT_{C-Tgt} - RTT_{C-S}$ then approximates the true
RTT between the SDN switch and probe target. Targets that do not respond to
\sami probes have no ICMP Echo Reply timestamp and leave us unable to calculate
the RTT; we therefore discard any targets with no probe replies. A REST API call
retrieves the timestamp values stored by \sami for analysis by the \sami client.

Because the ICMP ID field is 16 bits, this implementation allows for
approximately 65 thousand \texttt{ping} targets before \sami's state table is
full. We work around this limitation by pulling the current state from \sami
via a REST API call; specifically a GET request to a specific URL returns a JSON
array of \sami's state table to the \sami experimenter. Another REST API call
clears the state table and measurements can be resumed, thus allowing for a
potentially unlimited number of \texttt{ping} targets. An alternative
implementation might leverage the payload field of the ICMP Echos to maintain
the state of \texttt{ping} targets; we abstain from this approach in order to
maintain consistency of sent packet sizes for all probes when conducting
large-scale measurements.


\subsubsection{Traceroute}
\label{sec:traceroute}

Another common measurement task we implement in \sami is 
\texttt{traceroute}, initiated by a specific REST API
call\footnote{While there are several traceroute variants, we provide
basic implementation here to demonstrate the ease with which
measurements can be created.}. The
keywords contained in the JSON array of the HTTP PUT message are the target, and
number of probes per TTL.  \sami's \tr implementation operates by first
determining $RTT_{C-S}$ as described in Section~\ref{sec:calibration}. \sami
then creates ICMP Echo Request messages with TTL values beginning at 1,
delivering these to the switch via \pktout messages. Initially, the
ICMP Identifier value is set to 0, as in our \ping implementation. \sami
maintains TTL and target information for each probe by ICMP identification and
sequence numbers.  When intermediate routers respond with ICMP Time Exceeded
messages, \sami correlates the response with its corresponding
probe by parsing the ICMP quote,
and thereby determines the IP address and RTT for each hop along the
forward path. 
\sami increments the IP TTL value after each user-specified
number of probes at each TTL value, until one of two conditions occurs: i) the
\tr target is reached and responds with an ICMP Echo Reply, or ii) the TTL value
reaches 30 without having reached the \tr target. 
The \sami controller then
increments the ICMP Identifier value used for correlating messages bound for the
same \tr target together, as in our \ping implementation in
Section~\ref{sec:ping}. Our \tr implementation is
asynchronous; each successive ICMP packet is sent to the switch in a \pktout
message to be emitted as soon as it is created by the controller. \sami is
capable of handling these multiple packets in flight simultaneously due to the
state table it maintains of ICMP ID/Sequence Number values. Like our \ping
implementation, \tr probes for which intermediate routers or destination
addresses are unresponsive simply contain a null value as the return timestamp.
A REST API call returns the \sami \tr state to the \sami experimenter, who can
then reproduce the path from source to destination from the IP addresses.

\subsection{Custom Measurements}
\label{sec:custom}

In addition to performing existing active measurement tasks from
within SDN (\eg \ping and \tr as previously described), \sami
facilitates the creation and execution of new and novel measurements
in the spirit of network innovation.  As a proof-of-concept, we
partially implement the ``router ID'' primitive from our nascent work
in developing and using \emph{within} protocol
measurements~\cite{allman2016mip}.  

Today's method for active router discovery has remained essentially
unchanged for almost 30 years: \tr induces routers along the forward
path to a destination to return ICMP messages with one of the router's
interface IP addresses as the source.  Unfortunately, IP addresses are
a poor proxy for a router identifier.  Our reliance on \tr leads to
the cumbersome and error-prone processes of: i) alias resolution, to
determine the set of IP addresses belonging to the same physical
router; and ii) router ownership inference.  In the case of alias
resolution, the existing techniques require intensive active probing,
only work for a subset of addresses, and can both produce false
aliases and miss true aliases~\cite{keys10}.  Similarly, the \naive
method of using the autonomous system (AS) that originated the address
space to which a router's interface belongs is often not indicative of
the AS that owns or maintains the router, due to delegated and
off-path addresses~\cite{luckie2014second,bdrmap16}.

Our intent in implementing the router ID is not to perform exhaustive
measurements, but rather to provide a concrete example of the types of
measurement primitives that \emph{could} be created with \sami, and
their immediate benefit to network management and diagnosis.  

For debugging and management purposes, we imagine that a provider
wishes to extend the functionality of their core networking devices
such that they respond to an identification query with a unique router
identifier and the AS number to which the device
belongs.  Using \sami, we instantiate a rule in the switch to
encapsulate and forward to the controller any ICMP packets destined to
the switch with type 200 and code 0\footnote{corresponding to a
currently unused ICMP type}.  The controller parses this router ID
query message and creates a \pktout response with the device's AS and
identifier.  

This trivial example highlights several important characteristics of
\sami. First, while any database, \eg the DNS, could conceivably
provide an identical functionality, doing so requires keeping
different portions of the namespace consistently updated -- a
significant practical difficulty.  Instead, \sami provides a much
closer coupling where the control plane of the network device knows
its AS and router identifier.  Second, router ID demonstrates the ease
with which a new protocol or protocol extension can be implemented to
provide substantial measurement benefit.  In this case, router ID
effectively \emph{solves} both the aliasing and ownership problems
with an explicit mechanism, rather than the brittle and error-prone
inferences the measurement community is currently forced to employ.

\subsection{Deployment Scenarios}
We envision core network operators facilitating \sami experiments by allowing
researchers access to the \sami application running on their production
controllers. This immediately introduces several implementation challenges.
First, how might service providers arbitrate access to their \sami instance? Ryu
supports a robust PKI-based authentication system, allowing network operators to
permit \sami access to authorized users and specific switches via public key
authentication. Second, allowing \sami users the ability to inject an arbitrary
number of packets as quickly as the controller can generate them into their
network is likely an unappealing prospect for service providers. But because
\sami is a controller application, policies limiting maximum packet rates,
connection rates, and bitrates can be specified and tailored to individual
operators' risk assessments. Third, the potential exists for measurement
experiments to install flow rules on SDN switches that conflict with production
dataplane forwarding rules.  This potential conflict can be overcome by
assigning a lower priority to \sami user-generated flow rules, which would then
be ignored if a conflicting flow rule installed by the network operator exists,
thereby preventing unintended and potentially harmful consequences; \of already
supports this type of flow rule prioritization. Finally, \sami should support
multiple experimenters concurrently; because the controller is simply a
production server, partitioning of multiple users is a benefit inherited from
the deployment environment itself. Conflicting flow rules generated by
individual experimenters can be handled by assigning higher priority to certain
experiments, in a first-come, first-served manner, or according to
operator-specific instructions. Finally, to mitigate risk assumed by network
administrators running \sami instances, \sami can be configured to only allow
certain types of packets to be generated via its API. For instance, the ability
to send ICMP-based \ping or \tr probes at a low datarate, or packets to
TCP port 80 might be enabled, while the emission of more unusual packets (like
those discussed in Section~\ref{sec:custom}) could be restricted only to certain
users or disabled entirely.

\section{Results}
\label{sec:results}

\subsection{Ping}

\begin{figure}[t]
  \centering
  \resizebox{\columnwidth}{!}{\includegraphics{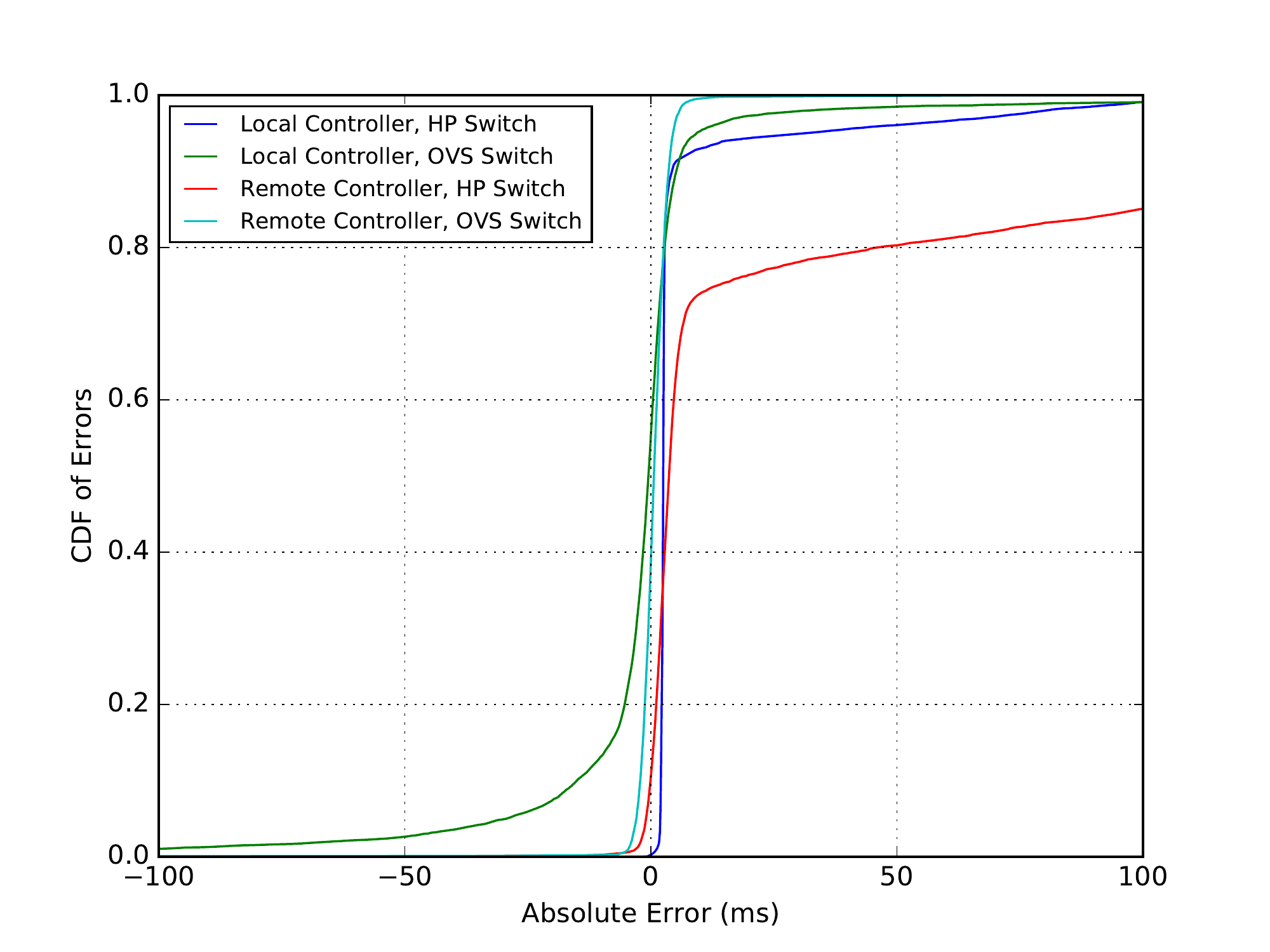}}
  \caption{Error between \sami-measured RTT and true RTT for $\sim$11,000
  \texttt{ping} targets (1 probe/target)}
  \label{fig:rttErrorCDF}
\end{figure}

\begin{figure}[t]
  \centering
  \resizebox{\columnwidth}{!}{\includegraphics{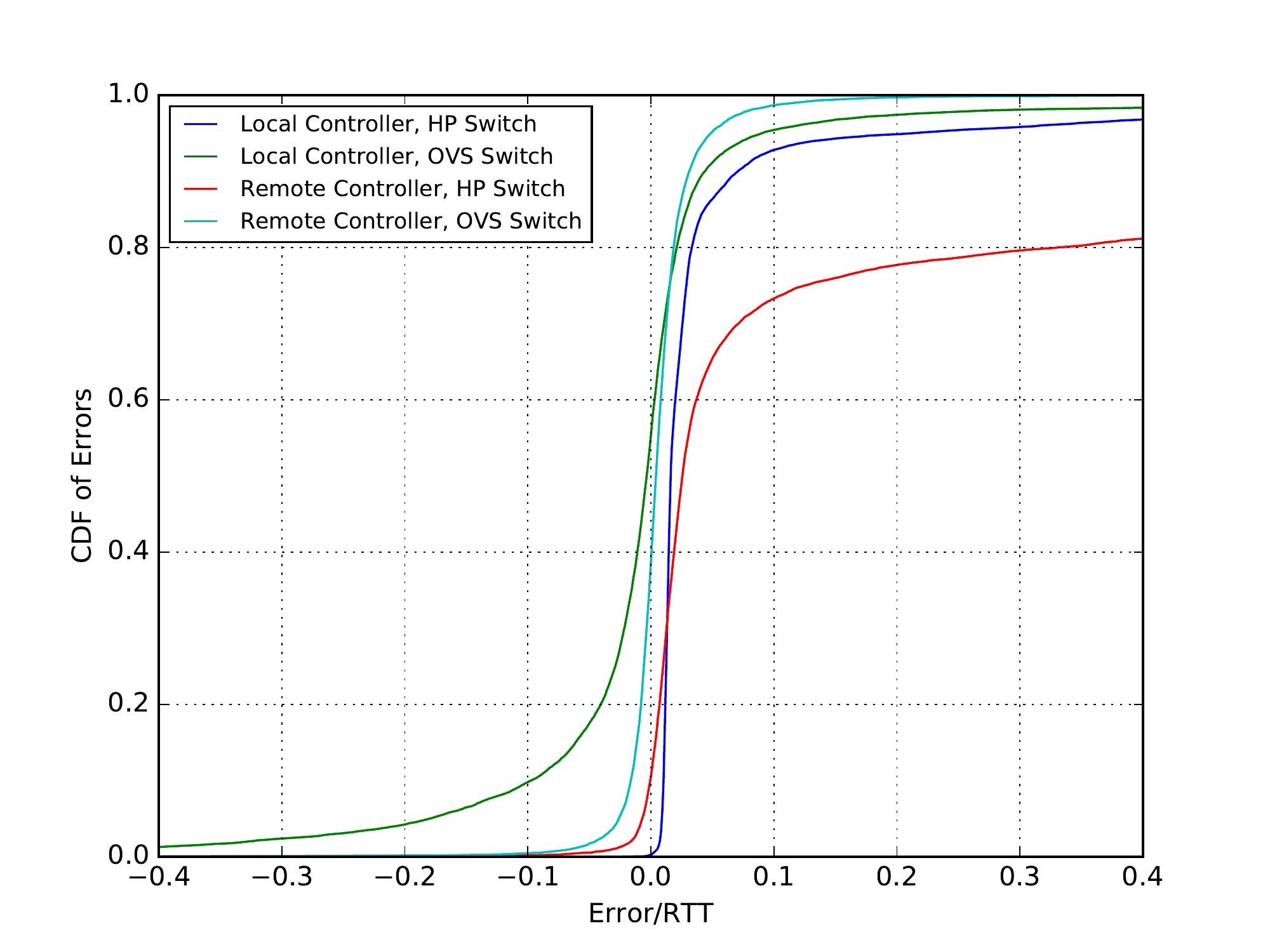}}
  \caption{\sami-measured error as a fraction of total RTT for \ping
  implementation.}
  \label{fig:rttErrorFractionCDF}
\end{figure}

\begin{figure}[t]
  \centering
  \resizebox{\columnwidth}{!}{\includegraphics{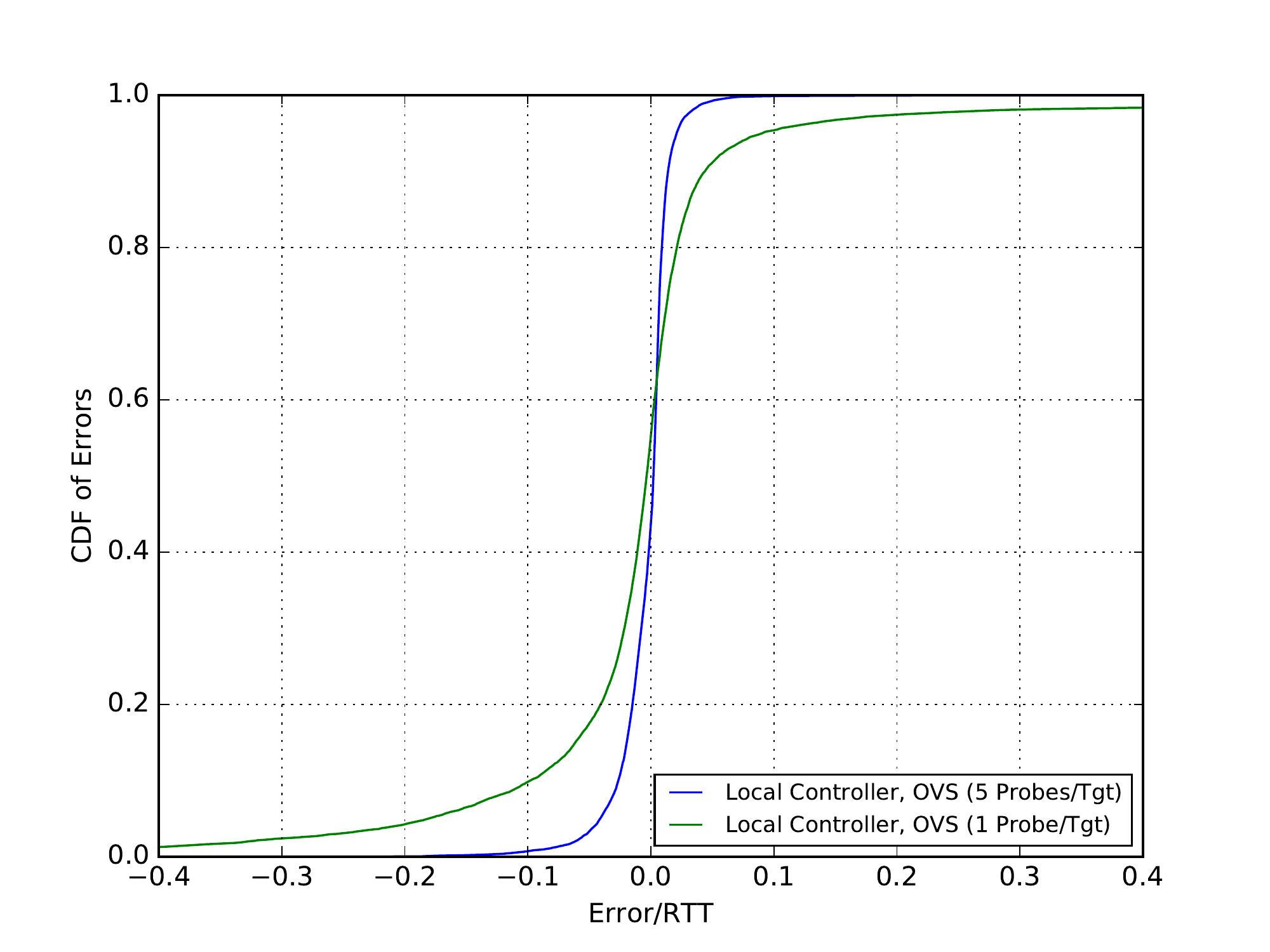}}
  \caption{Effect of additional probes on \sami-measured error from actual RTT
  using locally controlled \ovs switch}
  \label{fig:rttErrorMoreProbes}
\end{figure}

In order to measure the accuracy of \sami's \ping implementation, we capture
traffic entering and leaving the SDN switch through a hub connected to the
machine running \sami with \texttt{tcpdump}. In this manner, we 
obtain ground truth for the times our \sami-generated ICMP Echo Requests and
Replies exit and enter the switch. We account for various 
hardware and software implementations, using both an HP 2920 SDN switch and a
Linux machine running \ovs, as
well as remote versus local controller scenarios. Both the HP switch and machine
running \ovs are located in Monterey, California, while the remote
controller is located in Boston, Massachusetts. Figure~\ref{fig:rttErrorCDF}
shows the CDF of absolute error (between \sami-recorded RTT and actual RTT)
using an HP 2920 series switch, as well as using the machine running \sami as
the switch itself with \ovs (labeled as ``Local controller HP2920'' and
``Local Controller OVS'', respectively). Additionally, the software and hardware
switches were controlled from a \sami controller operating on the machine in
Boston, shown in the ``Remote Controller'' curves. All trials probed the same
15,000 IP addresses and received $\sim$11,000 replies each.  Our results show
that the \sami instance controlling the remote \ovs SDN switch has the
least amount of absolute error, with nearly 99\% of errors at 10 $ms$ or less.
The lower bound is the remote \sami instance controlling the HP switch, with
only $\sim$70\% of errors at the 10 $ms$ of less mark.
Figure~\ref{fig:rttErrorFractionCDF} is a CDF of the absolute error as a
fraction of the total RTT. In our \ping implementation, the \ovs
implementations consistently produce a lower $\frac{error}{RTT}$ value, with the
locally controlled \ovs trial obtaining an error of $<10\%$ of the total
RTT for more than 95\% of all probes. Both HP switch implementations produce the
most error in RTT measurement, with the remote controller architecture achieving
the same error percentage for approximately $20\%$ fewer RTTs. One method that
can be used to reduce the absolute and relative RTT errors is simply to send
more Echo Requests to the targets. In Figure~\ref{fig:rttErrorMoreProbes}, we
show the \sami error relative to the actual RTT in two trials, one in which five
Echo Requests were sent to the destination, and the other in which only one Echo
Request was emitted. For the trial in which five Echo Requests were sent to the
target, we record the mean RTT for all probes to each target. This strategy
dramatically reduces the overall RTT error.

\section{Related Work}
\label{sec:related}

Zeng \etal first posited the notion of SDN
controllers instructing switches to send test packets in the context
of their Automated Test Packet Generation system in
\cite{zeng2012automatic}.  We
extend this idea to the variety of active measurements in use today,
and implement such a system in real SDN hardware and software.

Our work relies on RTT measurements of probes to various network targets
generated and received by a controller from SDN switches; quantifying latencies
in SDNs is therefore fundamental to our work.
Rostos \etal develop the \emph{OFLOPS} platform
to evaluate \of switch implementations in various
use-cases \cite{rotsos2012oflops}.  OFLOPS examines processing delays in
\of switches when performing specific actions such as forwarding and
packet modification.  Our work also examines processing
delays in both hardware and software \of implementations of modern
devices. In
\cite{he2015latency}, He \etal measure inbound (switch to controller)
and outbound (controller to switch) latencies induced by \pktin events
and flow-modification rule insertions, deletions, and modifications,
respectively. Our timing analysis is impacted by delays caused by \pktin events
as well, which we similarly analyze on our devices using a passive tap.
\emph{SLAM}~\cite{yu2015software}, a tool used to monitor and estimate latencies
in data centers, sends packets between SDN switches to estimate the latency
along the path between them. \emph{SLAM} generates notifications to a controller
via \pktin notification messages sent by the first and last switches along a
path when a probe is received. As in our work, in \cite{yu2015software} the
authors account for controller to switch latency by continuously monitoring it
via \texttt{OFPEchoRequest} messages.

The notion of using network devices to perform active measurements
was first standardized in \cite{rfc4656} and \cite{rfc5357}, protocols
for performing one and two-way active measurements of delay and loss.
Our work is broader, generalizes these protocols, and provides a means for such
schemes to be implemented in network switches without explicit
vendor support.  
Most closely related to our own work is SDN
traceroute \cite{Agarwal:2014:STT:2620728.2620756}, a technique to 
reveal the sequence of switches and ports that actual data-plane
packets traverse in an SDN network.  SDN traceroute relies on
injecting measurement probes via \pktout messages, and installing rules to
match and retrieve tagged probe traffic via \pktin messages.  While
SDN traceroute and \sami rely on the same primitives, \sami is
designed to provide a platform for existing measurements, \eg IP-level
ping and traceroute, using SDNs.

\section{Conclusions and Future Work}
\label{sec:conclusions}


We introduce \sami, an active measurement application for SDNs. \sami enables
active measurement experiments to be conducted wherever SDNs are deployed,
bringing measurement capability to the core of the network as opposed to
traditional edge-based platforms. We implement two common active network
measurement utilities -- \texttt{ping}, to determine RTTs from an SDN switch
to network target, and an ICMP-based \texttt{traceroute}. To validate
our RTT measurements, we conduct multiple experiments to identify sources of
error and quantify their effects when using \sami as a measurement platform. We
find that these effects vary significantly according to the location of the
\sami controller and between hardware and software SDN switches, but can be
mitigated by sending several packets to each target, and using the mean RTT from
all probes. Finally, we demonstrate the ability to conduct custom measurements
by designing a capability to perform active router discovery by querying
routers for an imagined ``router ID'' value containing that device's ASN and an
identifying string. By sending an ICMP packet with an unused ICMP type and code
value to a router supporting this capability, \sami obtains topological data
about a device directly, rather than via current, error-prone
\texttt{traceroute} methods.

\subsection{Future Work}
With \sami instances deployed at the network core, we aim to understand forwarding
behavior not accessible to measurement experiments performed at the network
edge. For example, do \texttt{traceroute}s initiated from the network edge
follow the same terminal path to a destination as those initiated from a tier 1
AS? That is, do operator policies pertaining to source IP address affect the
path selection in ways that are not discernible to edge-based vantage points?
Additionally, as we note in Section~\ref{sec:custom}, SDN controllers are
capable of arbitrary packet creation. These packets need not adhere to the
constraints placed upon ordinary hosts; for example, a \sami controller may
generate IP datagrams with any source IP address. If destined for another \sami
switch, these spoofed source address packets can reveal source address
validation policies implemented by network operators unmeasurable by platforms
at the edge of the Internet. Finally, \sami's unique contribution of offering
measurement vantage points from the core of the network affords the opportunity
to conduct certain estimates, like bandwidth usage, closer to the quantity being
measured. We envision using \sami, or a similar system, to ``tag'' portions of
the traffic flowing through vantage points to better understand traffic
characteristics as it is routed through network operator-administered
infrastructure. 

Finally, while we focus on OpenFlow-based SDNs in this work due to the
ubiquity of their real-world deployment, we note that the SDN space is
rapidly evolving.  For example, recent work on programmable network
processors~\cite{bosshart2014p4} promises a much richer set of
abstractions and capabilities that \sami can utilize if realized.

\ifx\notdraft\true
\section*{Acknowledgments}
We thank Mark Allman, Steve Bauer, and Ethan Katz-Bassett for 
valuable early feedback.  
Views and conclusions are those of the authors and
should not be interpreted as representing the official policies or
position of the U.S.\ government.
\fi

\small{
\bibliographystyle{abbrv}
\bibliography{sami}
}

\end{document}